\documentclass[preprintnumbers,prd,twocolumn,showpacs,floatfix,preprintnumbers,letterpaper,amsmath,amssymb,nofootinbib,superscriptaddress]{revtex4}
\usepackage{graphicx}
\usepackage{epsfig}
\usepackage{bm}
\usepackage{amsfonts}
\usepackage{color}

\usepackage{amsmath,amssymb}

\begin{document}

\title{ Cosmology With  Axionic-quintessence Coupled with Dark Matter}

\author{Sumit Kumar\footnote{sumit@ctp-jamia.res.in}}
\affiliation{Centre for Theoretical Physics, Jamia Millia Islamia, New Delhi-110025, India}
\author{Sudhakar Panda\footnote{panda@hri.res.in}}
\affiliation{Harish-Chandra Research Institute, Chhatnag Road,
Jhunsi, Allahabad-211019, India}
\author{ Anjan A Sen\footnote{aasen@jmi.ac.in}}
\affiliation{Centre for Theoretical Physics, Jamia Millia Islamia, New Delhi-110025, India}

\date{\today}

\begin{abstract}
We study the possibility of explaining the late time acceleration with an axion field which is coupled with the dark matter sector of the energy budget of the Universe. The axion field arises from the Ramond-Ramond sector of the Type-IIB string theory. We study the background evolution of the Universe as well as the growth of the matter perturbation in the linear regime. We subsequently use the observational data from Sn-Ia, BAO measurements, measurements of the Hubble parameter as well as the observational data for the growth of the matter perturbation to constrain our model. Our results show that coupled axion models are allowed to have larger deviation from cosmological constant by the present observational data.
\end{abstract}

\pacs{98.80.Es,98.65.Dx,98.62.Sb}
\maketitle

\date{\today}

\maketitle

\section{Introduction}

Cosmological observations \cite{sn1, cmb, wmap, sdss} have now established the fact that our universe is undergoing a late time accelerating phase on large scales. This  has been confirmed by a number of precise cosmological observations.  One generally believes that an unknown component having a large negative pressure is driving this acceleration and efforts are on to explain this acceleration by suitable modification of gravity at large cosmological distances. For a detailed review on various model building aspects for this late time acceleration, we refer the readers to some excellent reviews on this subject \cite{review}.

If the acceleration is indeed driven by some unknown candidate, then a cosmological constant $\Lambda$  with an equation of state (e.o.s) $w=-1$  is the simplest option.  But the model is plagued by two very important theoretical issues: the fine tuning problem and the cosmic coincidence problem. Moreover, the observational data itself can allow models which are different from a constant $\Lambda$. These are termed as quintessence models \cite{quint}, and are primarily built with scalar fields with sufficiently flat potentials (Please see Copeland et al. \cite{copeland} for a detailed review on scalar field models). Problem with scalar field models are two folds: the energy scale involved with such fields is $\sim 10^{-3}$ eV, which is not only much less than the energy scale of inflation but also much less than that of the supersymmetry (SUSY) breaking.  Also, in order to achieve the slow-roll for this scalar fields ( which in turn ensure the negative e.o.s), the mass of these fields have to be of the order of $10^{-33}$ eV, and this should not be corrected to higher values due to SUSY breaking. This is extremely difficult to achieve given our understanding of the hierarchy problem in the standard model of particle physics.

 In a recent work,  Panda et al. \cite{pst} have tried to address these issues in the context of string theory (from now on PST model). The construction there is based on  the idea of axion monodromy \cite{axion} in the context of Type-II B string theory. The quintessence field in four dimensional spacetime was identified with a scalar field obtained by dimensional reduction of a rank two tensor field $C_{MN}$ coming from the Ramond-Ramond sector of the ten dimensional Type IIB string theory. This field is a tensor field and not a pseudo-tensor field in ten dimensions. The quintessence field is $C_{ab}$ (called axion) in four dimensions where both the indices $a$ and $b$ are along the compact directions and hence it is a scalar field in four spacetime dimensions and not a pseudo scalar. This is a parity even field not a parity-odd field like a standard axion. When the shift symmetry associated with this axion field is approximately broken, in a non-trivial way, the field develops a linear potential. One important thing that one should ensure is that the potential does not get corrections and is stable  up to the scale of supersymmetry breaking. In the PST model this has been done by choosing the field from the Ramond-Ramond sector since such fields do not couple to any other fields in the perturbative sector of the string theory ( which includes all the fields present in the Standard model). This ensures the flatness of the potential. Also as the quintessence field does not couple to the standard model field, this also helps to prevent the long range effect these quintessence field can have in our Solar system.

Given this attempt to construct a model for quintessence, in a subsequent work, the cosmological constraint on this model was derived by Gupta et al. \cite{gupta}, where it was showed  that the observational data (coming mainly from the background Universe) allow the model to behave very close to the $\Lambda$CDM model. This does not contradict the fact that a concordance $\Lambda$CDM is observationally most preferred model despite several unsolved theoretical issues.

As we discussed in the preceeding paragraphs, the quintessence field in this construction does not coupl to the standard model fields, that is the baryonic part of the matter energy density of the Universe. But the dark matter also contributes sufficiently to the energy budget of the Universe. In the string theory context, at present one does not have a microscopic theory for the dark matter and cetainly a lot has to be done in this respect. One may only speculate that the dark matter candidate(s) may come from the non-perturbative spectrum of the string theory, for examply D-particles, low level KK-modes of the Ramond-Ramond sector etc. It is also plausible to assume that the axion field, in the present context, can interact with the dark matter sector. This interaction could be also of non-perturbative nature. One obvious question is whether such coupling will spoil the flatness of the potential through the loop contribution. One can not say this at present. Due to our lack of understanding of the microscopic theory for the dark matter, we only consider dark matter energy density as a classical source in the Einstein's equations and hence will not enter in the loop calculations.  Once we understand the microscopic theory for dark matter better, this issue can be readdressed. 
 
In this paper, we investigate the cosmology with the PST-model of quintessence where the axion field is coupled with the dark matter sector of the theory but does not couple with visible matter sector of the Universe. We have already stated that previous investigations, where no such coupling exists, show that observational data constrain the PST model to behave very close to the $\Lambda$CDM Universe. Our motivation is to see whether allowing the coupling between the dark matter and the quintessencef ield can result in any significant deviation from this. We first study the background evolution in this set up a discuss the possible modifications that these coupling can result in the evolution of the Universe. We also study the evolution of the matter perturbation in linear theory. Finally we use different observational data to study the constraint on different parameters in our model.

In the next section, we formulate the equations for the background evolution of the Universe assuming the axion field is coupled to the dark matter sector of the Universe. In section III, we formulate the matter perturbation the linear regime in our model. We discuss the evolution of the linear growth of the matter fluctuations. In section IV, we discuss the observational constraints in our model using various observational data. Finally we conclude in section V.

\vspace{2mm}
\section{Background evolution with Coupled Axion}
\vspace{2mm}
We start with an interacting picture where the axion field is coupled with the dark matter sector (d) of the Universe. The visible matter sector (b) is not coupled with the axion field. As we discuss in the introduction, due to our lack of understanding the microscopic theory for dark matter, we can only allow the coupling between the dark matter and the dark energy sector in a phenomenological way. There have been number of investigations in the literature (named as "Coupled Quintessence") \cite{coupled} where such coupling has been introduced to study the cosmology. In our study, we follow the same set up together with the quintessence field in the PST model.

The relevant equations are given by

\begin{eqnarray}
 \ddot{\phi}+\frac{dV}{d\phi}+3H\dot{\phi}=C(\phi)\rho_{d} \nonumber \\ 
 \dot{\rho}_{d}+3H(\rho_{d})=-C(\phi)\rho_{d}\dot{\phi} \\
 \dot{\rho_b}+3H(\rho_b)=0 \nonumber\\
  H^2=\frac{\kappa^2}{3}(\rho_b+\rho_{d}+\rho_\phi) \nonumber \\
\end{eqnarray}

\noindent
together with the flatness condition
\begin{equation}
 1 = \frac{\kappa^2\rho_{b}}{3H^2}+\frac{\kappa^2\rho_{d}}{3H^2}+\frac{\kappa^2\dot{\phi}^2}{6H^2}+
 \frac{\kappa^2V(\phi)}{3H^2}
\end{equation}
 Here $C(\phi)$ represents coupling parameter between the axion field and dark matter. Due to our ignorance about the detail physics for this underlying interaction between the axion field and the dark matter, we address this issue at a phenomenological level and assume $C(\phi)$ to be a constant following the earlier work by Amendola \cite{coupled}. For $C=0$ we recover the uncoupled case.

\noindent
Next, we define the following dimensionless variables:
\begin{align}
 x	= \frac{\kappa\dot{\phi}}{\sqrt{6}H}, \hspace{1mm}
 y	= \frac{\kappa\sqrt{V(\phi)}}{\sqrt{3}H} \nonumber \\
 z	=\frac{\kappa\sqrt{\rho_b}}{\sqrt{3}H}, \hspace{1mm}
 \lambda	=	\frac{-1}{\kappa{V}}\frac{dV}{d\phi} \hspace{1mm}
 \Gamma	=	\frac{V\frac{d^2V}{d\phi^2}}{\left(\frac{dV}{d\phi}\right)^2} 
\end{align}

\begin{figure}
\includegraphics[width=2.6in,height=2in,angle=0]{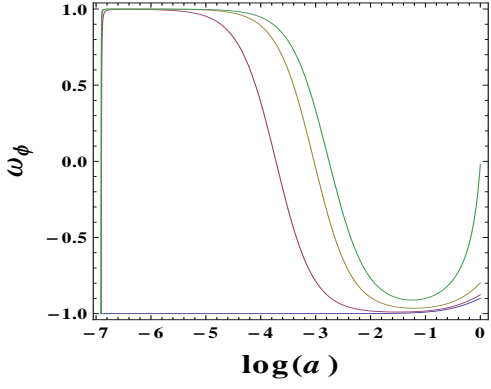}
\caption{Behaviour  of the equation of state for the axion field. Bottom to Top: $W=0, 0.01, 0.03,0.06$. $\Omega_{d0} = 0.23, \Omega_{b0} = 0.05, \lambda_{i} = 0.7$.}
\end{figure}

\begin{figure}
\includegraphics[width=2.6in,height=2in,angle=0]{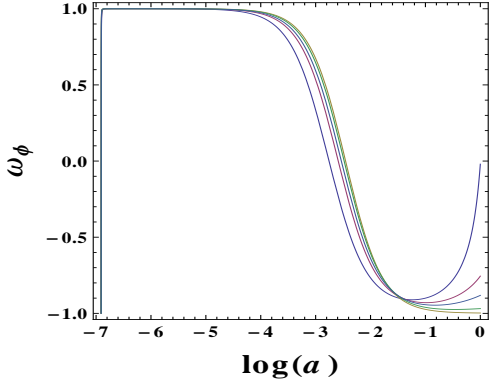}
\caption{Behaviour  of the equation of state for the axion field. Bottom to Top: $\lambda_{i}=0.05, 0.3, 0.5,0.6,0.7$. $\Omega_{d0} = 0.23, \Omega_{b0} = 0.05, W = 0.06$.}
\end{figure}

The density parameter $\Omega_{\phi}$ and the equation of state for the axion field $w_{\phi}$ are given as,
\begin{align}
 \Omega_\phi	=&	x^2+y^2 \\
 \gamma	=&	1+w_\phi	=	\frac{2x^2}{x^2+y^2}
\end{align}

\noindent
With this, one can form an autonomous system of equations:
\begin{align}
 \Omega_\phi'	=&	W\sqrt{3\gamma\Omega_\phi}(1-\Omega_\phi-z^2)+3\Omega_\phi(1-\Omega_\phi)(1-\gamma) \nonumber \\
 \gamma'	=&	W\sqrt{\frac{3\gamma}{\Omega_\phi}}(1-\Omega_\phi-z^2)(2-\gamma)+\lambda\sqrt{3\gamma\Omega_\phi}(2-\gamma)\nonumber\\
 - &3\gamma(2-\gamma) \nonumber \\
 z'	=&	-\frac{3}{2}z\Omega_\phi(1-\gamma) \nonumber \\
 \lambda'	=&	\sqrt{3\gamma\Omega_\phi}\lambda^2(1-\Gamma),
\end{align}
where $W=\frac{C}{\kappa}$. We evolve the above system of equations from the decoupling era ($a=10^{-3}$) to the present day ($a=1$). Given the  initial conditions for $\gamma$, $\Omega_{\phi}$ $z$ and $\lambda$, we can solve the system. 

The scalar field is initially  frozen due to large Hubble damping, and this fixes the initial condition for $\gamma$, $\gamma_{i} \approx 0$.  The initial value for $\lambda$, $\lambda_{i}$, is a parameter in our model.  $\lambda_{i}$ determines the deviation of $w_{\phi}$ from the initial $w_{\phi} \sim -1$  frozen state as the universe evolves with time. For smaller $\lambda_{i}$, the deviation extremely small, and the scalar field behaves as a cosmological constant at all time. For larger values of $\lambda_{i}$, $w_{\phi}$ has large deviation from $-1$ as the universe evolves.  In general, the contribution of scalar field to the total energy density  is negligible in the early universe. Nevertheless one has to fine tune the initial value of $\Omega_{\phi}$ in order to have its correct contribution at present. In this regard, $\Omega_{\phi}(initial)$ is related to the $\Omega_{\phi} (z = 0)$. Similarly the initial value of $z$ ( which is related to the density parameter for visible matter) can be related to the present day value of the $\Omega_{b}$ which is fixed at $\Omega_{b0} = 0.05$ in our subsequent calculations.
\begin{figure}[t]
\includegraphics[width=2.6in,height=2in,angle=0]{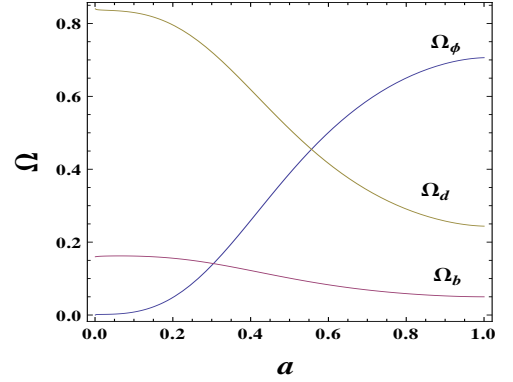}
\caption{Behaviour  of the density parameters for different components. $\Omega_{d0} = 0.23, \Omega_{b0} = 0.05, \lambda_{i} = 0.7, W = 0.06$.}
\end{figure}

\begin{figure}[t]
\includegraphics[width=2.6in,height=2in,angle=0]{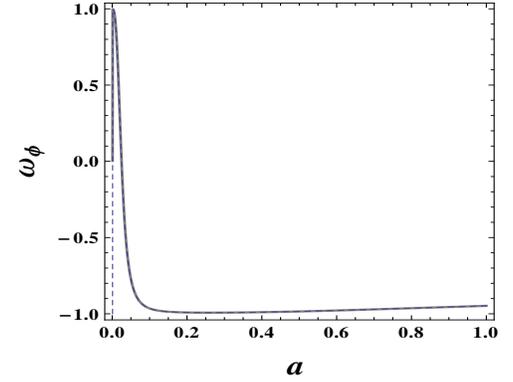}
\caption{Behaviour  of the equation of state for the axion field.$W=0.01$, $\lambda_{i}=0.5$. $\Omega_{d0} = 0.23, \Omega_{b0} = 0.05$. The solid line is for $\gamma_{i}=1$ and the dashed line is for $\gamma_{i} \sim 0$.}
\end{figure}
In our study the scalar field is the axion field originated from the Ramond-Ramond sector of the Type-IIB string theory. The form of the potential for this field is given by
\begin{equation}
V(\phi) = \frac{\mu^4}{f_{a}}\phi,
\end{equation}
where
\begin{equation}
f_{a}^2 = \frac{g_{s}^2 M_{pl}^2}{6 L^4}, \hspace{1mm} \mu^4 = \mu_{1} + \mu_{2}.
\end{equation}
Here $\mu_{1} = \frac{2 e^{4 A_0}}{(2\pi)^5 g_s \alpha'^{2}}$ and  $\mu_{2} = c M_{SB}^4 e^{2 A_0} \left( \frac{R^2}{{\alpha' } L^4}\right)$. $g_{s}$ is the string coupling constant and $V_{1} = L^{6}\alpha^{'3}$ is the volume of the internal space in the ten dimensional space time. $A_{0}$ is related to the warp factor at the location of NS5 brane and $R$ is the radius of the Ads like throat ( See \cite{pst}  for the detail construction of this model). It can be shown that the $\mu_{2}$ term have the dominant contribution in the potential $V(\phi)$. With this form of the potential, $\Gamma = 0$  in the system of equations (7). 
\begin{figure}[t]
\includegraphics[width=2.6in,height=2in,angle=0]{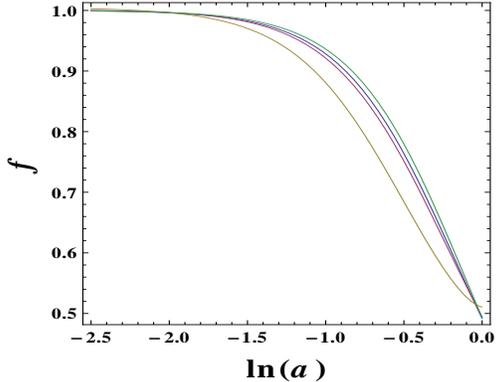}
\caption{Behaviour of the growth factor $f=\frac{d (log \delta_{t})}{d (log a)}$  as a function of the scale factor. From top to bottom, $\Lambda$CDM, $W=0, 0.03,  0.06$. $\Omega_{d0}=0.23$ and $\Omega_{b0} = 0.05$. $\lambda_{i} = 0.7$ for models different from $\Lambda$CDM.}
\end{figure}
In figure 1 and 2, we show the behaviour of the equation of state for the axion field $w_{\phi}$ as a function of redshift keeping $\lambda_{i}$ fixed  and varying $W$ ( figure 1) and keeping $W$ fixed at varying $\lambda_{i}$ ( figure 2). From these figures it is clear that $w_{\phi}$ around present epoch ($a \sim 1$) deviates more from $w=-1$ for larger $\lambda_{i}$ as well as for larger $W$.  For smaller values of these two parameters,  $w_{\phi}$ behaves very close to the $\Lambda$ value $w=-1$. In figure 3, we show the behaviour of the density parameters for different components as a function of scale factor. It is evident that in the early era, Universe is mostly dominated by the visible as well as dark matter components. But as the Universe evolves, the axion field starts dominating in the late epoch. 
\begin{center}
\begin{figure*}[t]
\begin{tabular}{c@{\qquad}c}
\epsfig{file=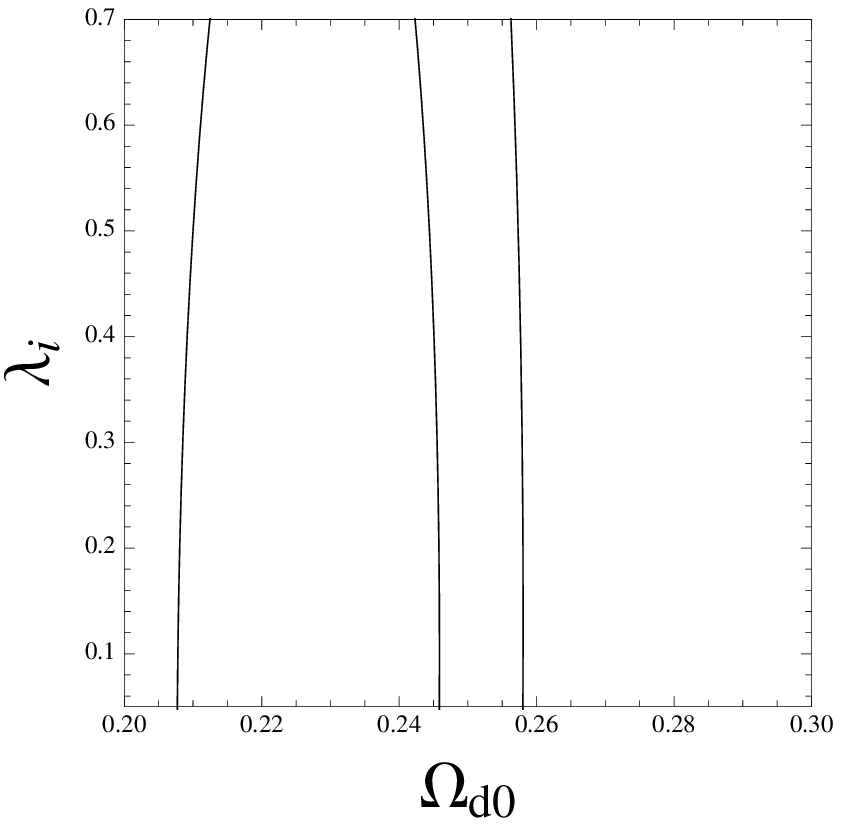,width=6 cm}&
\epsfig{file=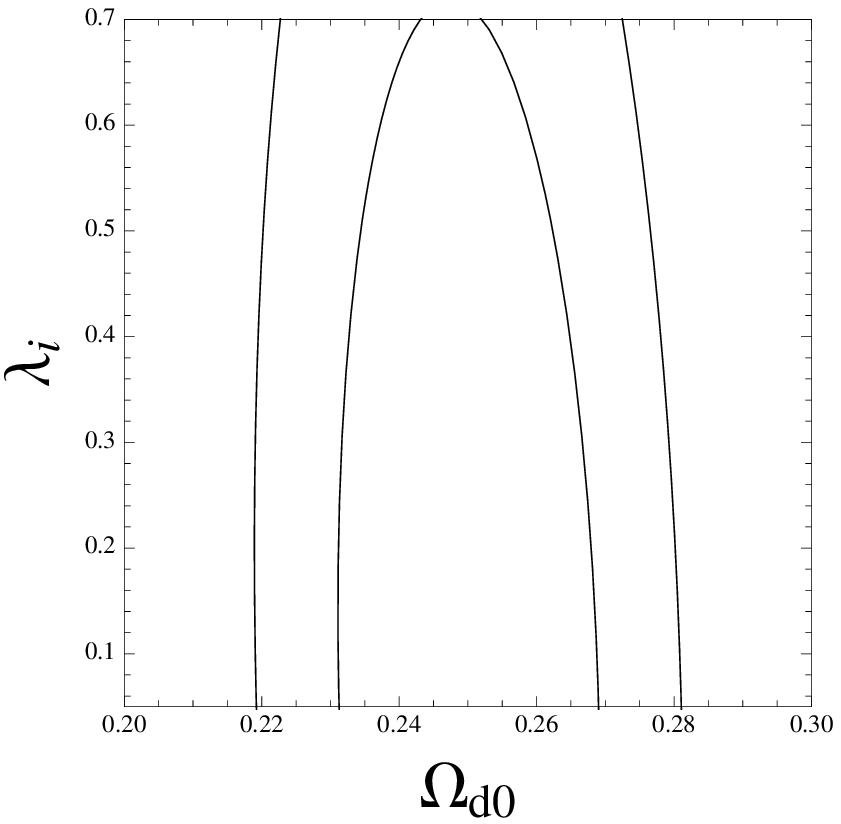,width=6 cm}\\
\epsfig{file=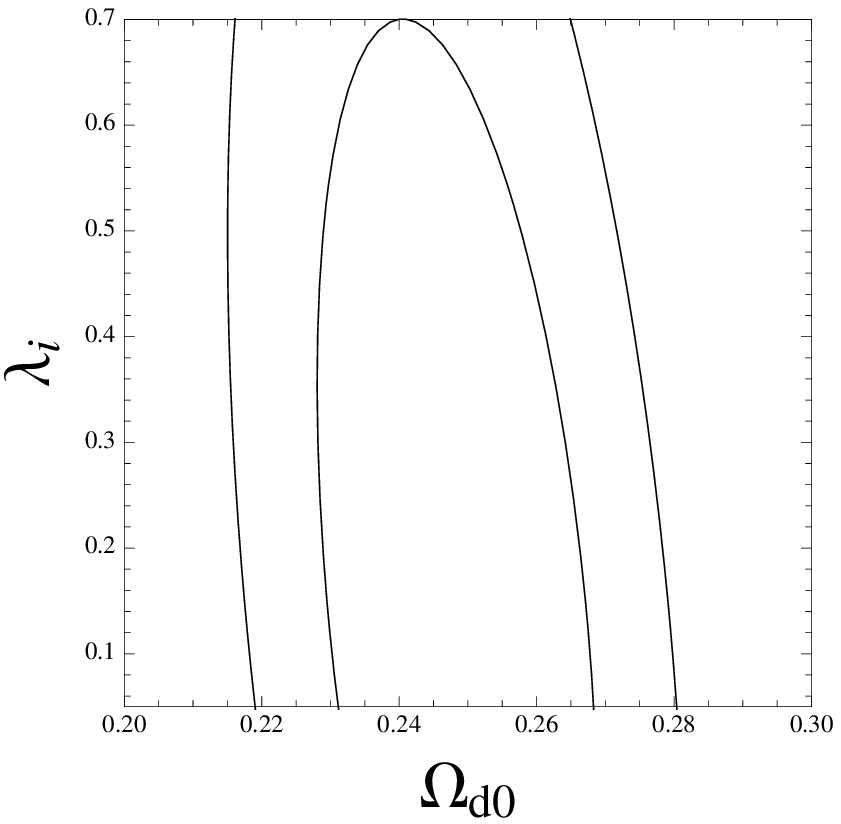,width=6 cm}&
\epsfig{file=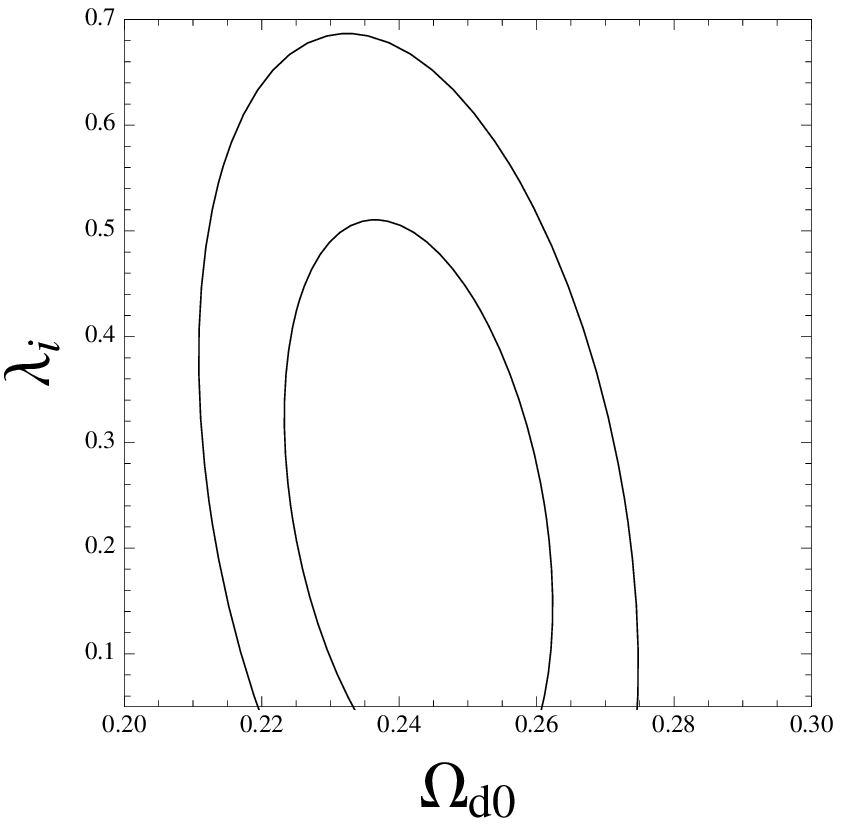,width=6 cm}\\
\end{tabular}
\caption{$68\%$ and $95\%$ confidence contours in $\Omega_{d0}-\lambda_{i}$ plane. Top Left: uncoupled case $W=0$, Top Right: coupled case $W=0.01$, Bottom Left: coupled case $W=0.03$, Bottom Right: coupled case $W=0.06$. }
\end{figure*}
\end{center}

 We mention that the axion field is initially frozen due to large Hubble damping and that sets the initial $w_{\phi i} = -1$.  But there is a jump initially in the $w_{\phi}$ from this frozen state as can be seen from the figures 1 and 2. This is due to the energy transfer due to the coupling between the axion field and dark matter component. This coupling depends on the dark matter energy density which is large during decoupling time $(a \sim 10^{-3}$). Hence even if the potential is sufficiently flat and does not contribute much to the rolling of the field, the energy transfer through coupling, helps the scalar field to fast roll and the equation of state quickly reaches the stiff equation of state ($w=1$). The scalar field continues to remain in this fast roll phase till the Hubble damping starts dominating again. It Subsequently slows down the fast roll phase and the scalar field is frozen again. There after it remains frozen until the Hubble damping decreases sufficiently so that the scalar field slowly thaws away from this frozen state. The initial jump in equation of state from the frozen state and its subsequent approach to another frozen state crucially depends on the interplay between the energy transfer due to coupling and the large Hubble damping. It is also important to note that this initial jump in the equation of state of the scalar field does not affect the background evolution of the Universe as its contribution to the energy budget of the Universe during this period is sufficiently small ( See figure 3).  We should mention that the choice of the initial frozen state ($\gamma_{i} \sim 0$) does not play significant role in the scalar field evolution. Even if we assume that initially the scalar field mimics the background matter fluid ($\gamma_{i} \sim 1$), the evolution would have been exactly the same.  We show this in figure 4, where we plot the time evolution of equation of state for the  the scalar field in the coupled case, with two initial conditions, $\gamma_{i} \sim 0 $ and $\gamma_{i} \sim 1 $, keeping the other model parameters the same. It shows that the late time behaviours are exactly the same for the two cases. We should also mention this overall evolution of the scalar field remains the same even if we set our initial condition prior to decoupling era. 

The initial stiff jump in the equation of state does not create any pathology in Universe's evolution; because during the period when this sudden jump happens the contribution from the axion field to the total energy budget of the Universe is negligible as can be seen from figure 3. So the axion field does not affect the Universe's evolution during this time. We should point that similar things happens when a thawing scalar field is coupled to the total matter component (visible + dark matter) of the Universe \cite{gav}.
\vspace{2mm}
\section{Linear Perturbation}
\vspace{2mm}

\noindent
In this section, we write down the equations that governs the growth of fluctuations for dark matter and visible matter. In our model the dark matter is coupled to the axion field whereas the the visible matter is uncoupled. We work in the longitudinal gauge,
\begin{equation}
ds^{2} = a^{2}\left[-(1+2\Phi)d\tau^2 + (1-2\Psi) dx^{i}dx_{i}\right],
\end{equation}

where $\tau$ is the conformal time and $\Phi$ and $\Psi$ are the two gravitational potential. It is well known that in the absence of any anisotropic stress $\Phi= \Psi$.
We follow the prescription given by Amendola \cite{coupled} and write the equations for the density perturbations for visible matter and dark matter in the Newtonian limit which is valid for small scales. In this scale, one can safely assume that the axion field does not cluster otherwise it will behave like a massive dark matter. With these assumptions, the equations governing the growth of the linearised fluctuations in dark matter and visible matter  are given by:
\begin{equation}
 \delta''_{d}+\left(1+\frac{\mathcal{H}'}{\mathcal{H}}-2\beta_{d}x\right)\delta'_{d}-\frac{3}{2}(\gamma_{dd}\delta_{d}
 \Omega_{d} + \gamma_{db}\delta_{b}\Omega_{b})=0
\end{equation}
\begin{equation}
 \delta''_{b}+\left(1+\frac{\mathcal{H}'}{\mathcal{H}}-2\beta_{b}x\right)\delta'_{b}-\frac{3}{2}(\gamma_{db}
 \delta_{d}\Omega_d+\gamma_{bb}\delta_{b}\Omega_{b})=0
\end{equation}

\noindent
Here $\gamma_{ij}=1+2\beta_{i}\beta_{j}$ and $\mathcal{H}$ is conformal Hubble parameter $\mathcal{H}=aH$. For our case, $\beta_b=0$ and $\beta_d=W$. The density fluctuations $\delta_{d}$ and $\delta_{b}$ are defined as 
$\delta_{i} = \frac{\delta\rho_{i}}{\rho_{i}} $.  The total density fluctuation for the matter component ( dark + visible) is given by
\begin{equation}
\delta_{t} = \frac{\delta_{d}\Omega_{d} + \delta_{b}\Omega_{b}}{\Omega_{d}+\Omega_{b}}.
\end{equation}

\noindent
In figure 4, we show the behaviour of the growth function $f = \frac{d (log \delta_{t})}{d (log a)}$ as a function of scale factor. One can see that the growth is suppressed as one increases the coupling constant $W$.
\vspace{2mm}
\section{Observational Constraints}
\begin{center}
\begin{figure*}[t]
\begin{tabular}{c@{\qquad}c}
\epsfig{file=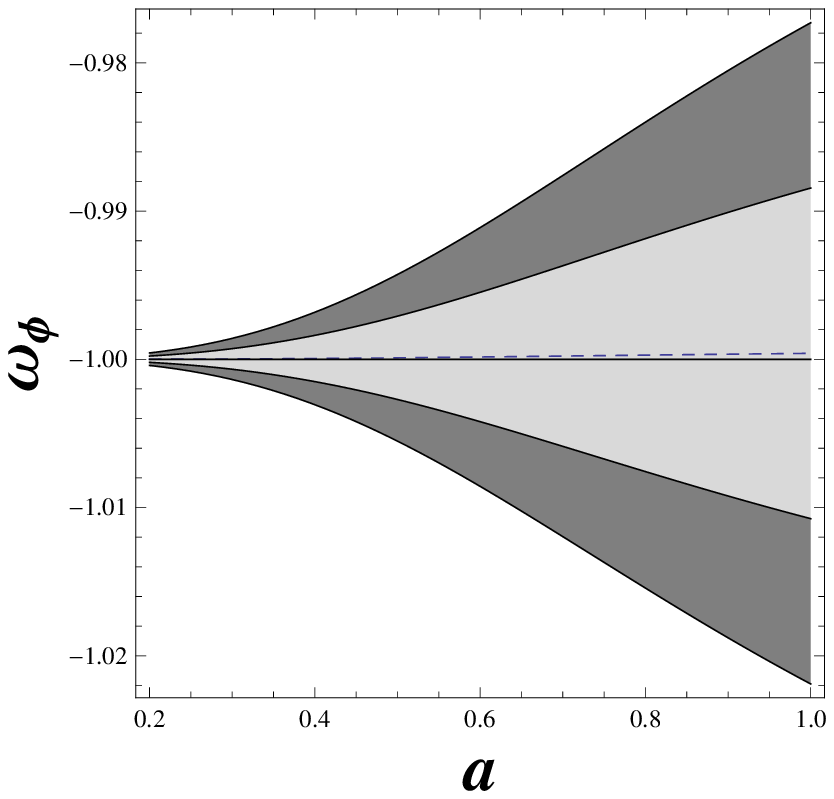,width=6 cm}&
\epsfig{file=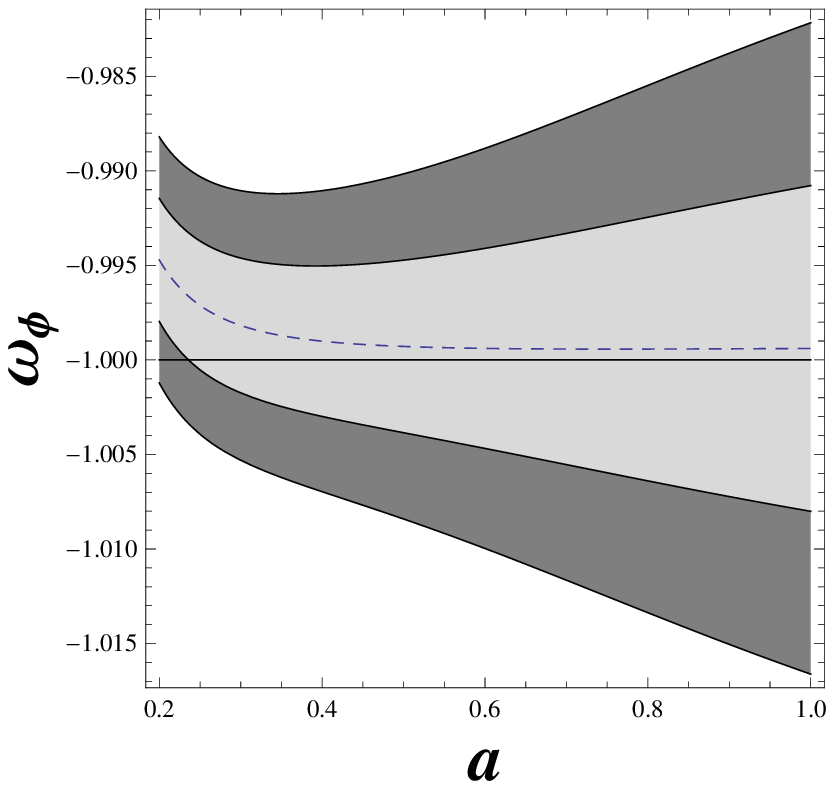,width=6 cm}\\
\epsfig{file=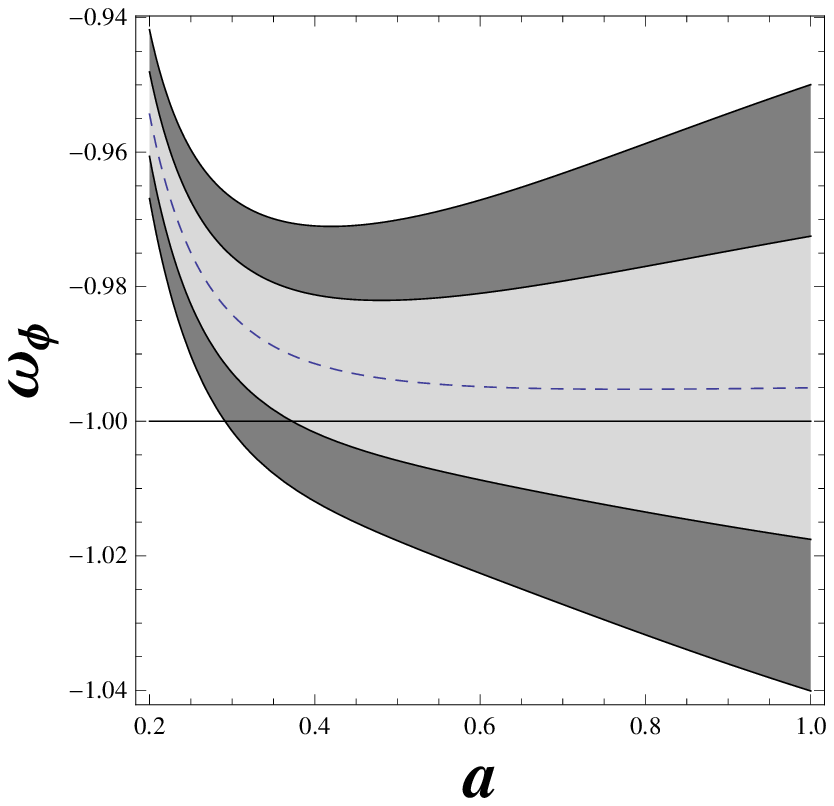,width=6 cm}&
\epsfig{file=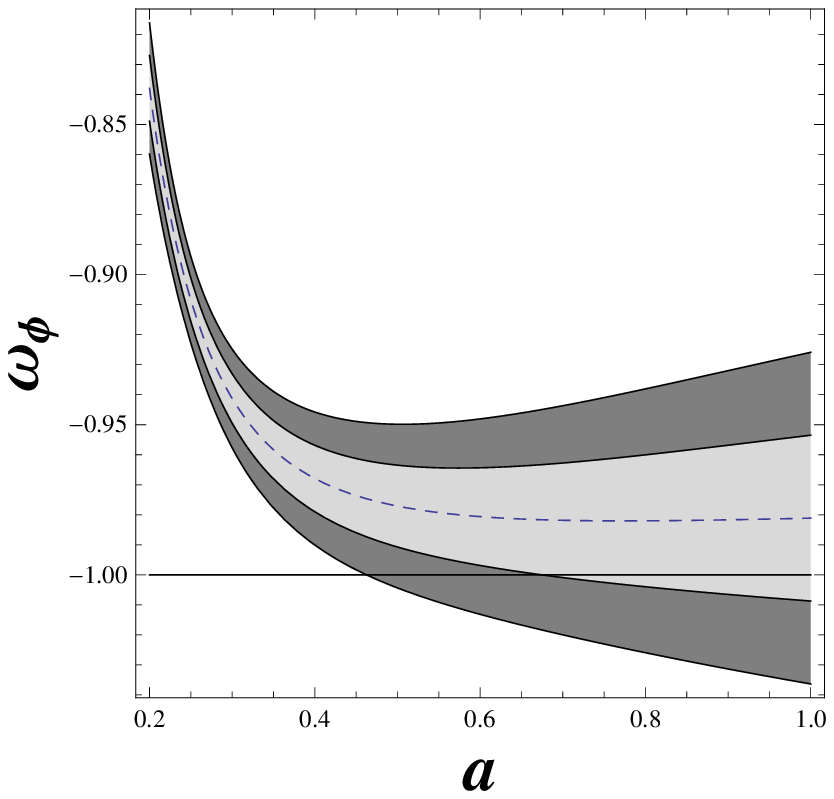,width=6 cm}\\
\end{tabular}
\caption{Allowed behaviour for the equation of state of the axion field $w_{\phi}$ at $68\%$ and $95\%$ confidence  level. Top Left: uncoupled case $W=0$, Top Right: coupled case $W=0.01$, Bottom Left: coupled case $W=0.03$, Bottom Right: coupled case $W=0.06$. Light Shade for $68\%$ and dark shade for $95\%$. The dashed line represents the best fitted behaviour and the solid line represents line represents the phantom divide.}
\end{figure*}
\end{center}
\vspace{1mm}
\noindent
In this section, we use the current observational data to put constraints on our  model (see \cite{coupquint} for earlier works on observational constraints for coupled quintessence).  We start with the Supernova Type-Ia observations which  directly probes the cosmological expansion. It actually measures the luminosity distance of different supernova explosions at different redshifts which is defined as:
\begin{equation}
d_{L}(z) = (1+z) \int^{z}_{0}\frac{1}{H(z')}dz'.
\end{equation}
The actual observable, {\it distance modulus} is given by
\begin{equation}
\mu = 5 log \frac{d_{L}}{Mpc} + 25.
\end{equation}
We consider the Union2.1 compilation containing 580 data points for $\mu$ at different redshifts \cite{suzuki}.

Next, we use the observational constraints on Hubble parameter which have been recently compiled by Moresco et al. \cite{Moresco2012}  in the redshift range $0 < z < 1.75$ using the differential evolution of the cosmic chronometers.  They have built a sample of 18 observational data point for $H(z)$ spanning almost $10$ Gyr of cosmic evolution. 

Next, we use the combined BAO/CMB constraints as derived recently by Giostri et al. \cite{giostri12}. This gives the constraint on the angular scales for the Baryon Acoustic Oscillation peak in the matter power spectrum. At present we have measurements from SDSS survey \cite{Percival} , 6dF Galaxy Survey \cite{beutler} and more recently, by the WiggleZ team \cite{blake11}. The corresponding covariance matrix is given by Giostri et al \cite{giostri12}.

We further use the measurement for the growth parameter $f = \frac{d log \delta_{t}}{d \log a}$.  The list of all the currently available growth data is given in the reference \cite{somgav} ( see also references therein).


With this, we calculate the combined likelihood:
\begin{equation}
-2 \log {\cal L} = - 2 \log ({\cal L}_{sn} \times {\cal L}_{bao} \times {\cal L}_{hub} \times {\cal L}_{gr}).
\end{equation}.

\noindent
This likelihood function is a function of the model parameters $\Omega_{d0}, \lambda_{i}$ and $W$. As we mentioned earlier, we have fixed $\Omega_{b} = 0.05$ for the visible matter. We maximize the likelihood function for $W=0$ for the uncoupled case and $W=0.01, 0.03, 0.06$ for the coupled case and draw the corresponding confidence contours in the $\Omega_{d0}-\lambda_{i}$ plane where $\Omega_{m0} = \Omega_{d0}+\Omega_{b0}$.

\noindent
In figure 6, we show the confidence contours in the $\Omega_{d0}-\lambda_{i}$ plane  for different values of the coupling constant $W$. We fix $\Omega_{b0}=0.05$ for our calculations. The first thing to be noticed is that, for the uncoupled case ($W=0$), $\Omega_{d0}$ and $\lambda_{i}$ are completely uncorrelated. As one increases the value of $W$, the two parameters become slightly anti correlated. One concludes from these plots that in the coupled case the upper bound on the initial slope of the axion potential $\lambda_{i}$ is comparatively smaller than the uncoupled case. As the value of the axion field is inversely proportional to the slope of the potential in our model, this translates to a bigger lower bound for the axion field in the coupled case.

 We have discussed in the earlier section that the deviation of the cosmological evolution from the concordance $\Lambda$CDM behaviour is controlled by the two parameters $\lambda_{i}$ and $W$.  Hence we use the standard error propagation technique \cite{coe} to calculate the error in the equation of state $w_{\phi}$ knowing the errors in $\lambda_{i}$ and $\Omega_{d0}$ for different values of $W$. The results are shown in the figure 7. In these figures, we also show the region in the phantom side as obtained by the error propagation. But for our axion field the relevant region is above the line $w_{\phi} =-1$. We observe from these plots that for the uncoupled case, the allowed behaviour for $w_{\phi}$ is very close to the $\Lambda$ line $w=-1$. This is consistent with the earlier results obtained by Gupta et al. \cite{gupta}. As one increases the coupling parameter $W$, larger deviation from $w=-1$ is allowed. This confirms that the coupling with dark matter can result the axion-quintessence field to behave differently from the cosmological constant $\Lambda$. 
 
 \begin{figure}
\includegraphics[width=2.7in,height=2.1in,angle=0]{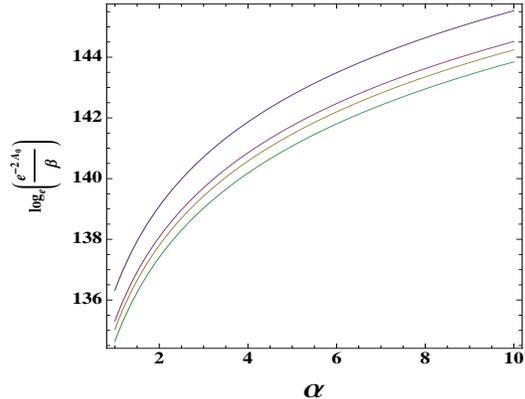}
\caption{The allowed value for $\left[\frac{e^{-2A_{0}}}{\beta}\right]$ as a function of SUSY breaking scale $\alpha$ (in TeV) assuming the best fit values for the cosmological parameters. From top to bottom, $W= 0, 0.03, 0.04, 0.06$.}
\end{figure}
 Next we calculate the constraints on the parameters that appear in the expression for the potential given in equation (8). We assume that the dominant contribution comes from the $\mu_{2}$ term \cite{pst,gupta}. After straightforward algebraic calculations, one can write
 \begin{equation}
 M_{SB}^{4} \approx M_{pl}^{2} H_{0}^{2} \left(\frac{e^{-2A_{0}}}{\beta}\right)s,
 \end{equation}
 where $\beta = \frac{R^2}{g_s\alpha'L^2}>1$, $s= (\frac{\mu^4}{f_a})/(\sqrt{3}M_{pl}H_{0}^{2})$. We take $H_0\sim 10^{-42}$~GeV and $M_{pl}\sim 10^{19}$~GeV and we assume $M_{SB} = \alpha$ Tev.

\noindent
Using the best fit values for the different cosmological parameters, one can easily get the corresponding value for $s$. From this, either one can assume a value for the combination $\left[\frac{e^{-2A_{0}}}{\beta}\right]$ and a get a bound on the SUSY breaking scale $\alpha$, or assume a value for $\alpha$ and get a bound on the combination $\left[\frac{e^{-2A_{0}}}{\beta}\right]$. We show this in figure 8.

\section{Conclusion}
The PST model proposed by Panda et al. is one interesting attempt to build  a model for quintessence in the context of string theory using the idea of axion monodromy. In this set up, the axion field arises in the Ramond-Ramond sector of string theory avoiding the interaction with standard model fields. But such an axion field can, in principal, interact with the dark matter sector of the Universe which can be nonperturbative in nature.

In this work, we study the cosmology in a PST model where the axion field is coupled with dark matter sector but not with the visible matter sector. Due to the lack of understanding about the detail physics for this interaction, we assume the coupling function to be a constant for simplicity following the various approaches for the couple quintessence models \cite{coupled}. 

 In our model, we evolve our system from the decoupling era upto the present day. There is always a stiff jump in the equation of state towards the stiff nature ($w_{\phi} = 1$) due to the energy transfer through coupling but subsequently the field is driven towards a frozen  state  ($w_{\phi} = -1$)  due to large Hubble damping in the early Universe. Later on, as the Hubble damping relaxes, the field slowly thaws away from this frozen state and deviates from the $w_{\phi} = -1$ depening upon the initial slope of the potential. Interestingly this behaviour remains the same whether one assumes $\gamma_{i} \sim 0$ (cosmological constant) or $\gamma_{i} = 1$ (background matter) for the scalar field.

We use the latest data from Supernova Type-Ia, Baryon acoustic oscillations, measurements of the Hubble parameter as well as the measurement of the growth, to constrain the model parameters. For the uncoupled case, we confirm the earlier results by Gupta et al.\cite{gupta} that data allow extremely small deviation from the cosmological constant $\Lambda$. But as one introduces the coupling between the axion field and the dark matter, larger deviation from the $w=-1$ behaviour is allowed with increasing strength of the coupling.

 Lastly, the other useful way to constrain this model further is by using the CMB large scale anisotropies and also through its cross-correlations with matter power spectra. The Inergrated-Sachs-Wolfe (ISW) effect will be quite prominent in this model and using the recent Planck results \cite{planck}, one can put more stringent constraints on this model. This will be our future goal. 

\section{Acknowledgement}

The author SK is funded by the University Grants Commission, Govt.of India through the Senior Research Fellowship. The author AAS acknowledges the funding from SERC, Dept. of Science and Technology, Govt. of India through the research project SR/S2/HEP-43/2009.

\end{document}